\documentclass[journal]{IEEEtran}
\usepackage{etex}
\usepackage{amsmath,amsfonts,amssymb}
\usepackage{pstricks}
\usepackage{graphicx}
\usepackage{cite}
\usepackage{epstopdf}
\usepackage{soul}
\usepackage{array}
\usepackage{subcaption}
\usepackage{tikz}
\usepackage{xcolor,colortbl}
\usepackage{stfloats}
\usepackage{todonotes}

\usetikzlibrary{calc,trees,positioning,arrows,chains,shapes.geometric,decorations.pathreplacing,decorations.pathmorphing,shapes,matrix,shapes.symbols,fit,backgrounds}
\tikzset{
>=stealth',
  punktchain/.style={
    rectangle,
    rounded corners,
    draw=black, very thick,
    text width=12em,
    minimum height=2em,
    text centered,
    on chain},
  line/.style={draw, thick, <-},
  every join/.style={->, thick,shorten >=1pt},
  decoration={brace},
  tuborg/.style={decorate},
  tubnode/.style={midway, right=2pt},
}

\definecolor{Gray}{gray}{0.90}

\usepackage{booktabs}

\DeclareMathOperator{\tr}{tr}

\DeclareMathOperator{\E}{\mathrm{E}}
\DeclareMathOperator{\Var}{\mathrm{Var}}

\newcommand{\MatVec}[1]{\boldsymbol{\mathsf{#1}}}

\newcommand{\RNum}[1]{\uppercase\expandafter{\romannumeral #1\relax}}

\usepackage{./picins/picins}

\newcolumntype{o}{>{\columncolor{white}}c}
\newcolumntype{e}{>{\columncolor{white}}c}

\begin{document}
\title{Aperiodic Array Synthesis\\for Multi-User MIMO Applications}
\author{C.~Bencivenni~\IEEEmembership{Member,~IEEE},
	A.~A.~Glazunov~\IEEEmembership{Senior Member,~IEEE},
	R.~Maaskant,~\IEEEmembership{Senior Member,~IEEE},
	and~M.~V.~Ivashina,~\IEEEmembership{Senior Member,~IEEE}
\thanks{C.~Bencivenni, A.~A.~Glazunov, R.~Maaskant, and M.~V.~Ivashina are with the Signals and Systems Department of the Chalmers University of Technology, G\"oteborg, Sweden; R.~Maaskant is also with the Electromagnetics Group of the Eindhoven University of Technology, the Netherlands, e-mail: carlo.bencivenni@chalmers.se, andres.glazunov@chalmers.se, rob.maaskant@chalmers.se/r.maaskant@tue.nl, marianna.ivashina@chalmers.se.}
\thanks{This work is financed by the Swedish VINNOVA Excellence Research Center Chase/ChaseOn, and by part by the Netherlands Institute for Scientific Research NWO VIDI grants.}
\thanks{Manuscript submitted to \emph{IEEE Trans. Antennas Propag.} on March 18, 2017; published and defended as part of the PhD dissertation \emph{Aperiodic Array Synthesis for Telecommunications} on May 31, 2017, G\"oteborg, Sweden, \emph{candidate}: Carlo Bencivenni, Chalmers University of Technology, \emph{faculty opponent:} Andrea Massa, ELEDIA center.}}
\maketitle
\begin{abstract}
This paper demonstrates the advantages of aperiodic arrays in multi-user multiple-input multiple-output systems for future mobile communication applications.
We propose a novel aperiodic array synthesis method which account for the statistics of the propagation channel and the adaptive beamforming algorithm.
Clear performance gains in line-of-sight dominated propagation environments are achieved in terms of the signal-to-interference-plus-noise ratio, the sum rate capacity, as well as the spread of the amplifier output power as compared to their regular counterparts. We also show that the performance is not sacrificed in rich scattering environments. Hence, aperiodic array layouts can provide performance gains in millimeter-wave applications with a dominating line-of-sight component.
\end{abstract}
\IEEEpeerreviewmaketitle
\section{Introduction}\label{Intro}
The continuously growing need for higher capacity and user data rates in wireless communications calls for new multi-antenna concepts, such as the massive multiple-input multiple-output (MIMO) concept.
The practical implementation of such complex antenna systems is very challenging, particularly if power-efficient and cost-effective solutions are to be realized. Typical systems require hundreds up to thousands of active antenna elements, each of which is equipped with a signal digitizing circuit~\cite{Rusek_MIMO_Massive:2013}. 

Research on massive MIMO solutions has mainly focused on classical uniform array layouts.
However, aperiodic array layouts are potentially advantageous in suppressing spatially distributed interference through minimizing side-lobe levels while maximizing the power efficiency of amplifiers through the use of isophoric array architectures. These advantages have been exploited in cases where the desired beamshape is known a priori, including satellite communication and radio astronomy applications~\cite{THIN_L1_Carlo:2016,APERIODIC_ASTRONOMY_Cappellen:2006}.
In MIMO systems, adaptive beamforming is employed, thus there is no a priori knowledge on the desired beamshape nor the element excitations, since these are both channel-state dependent and dynamically adapted to improve the link quality and/or capacity. Most of the available aperiodic array synthesis methods are therefore not suitable or readily applicable to such beamforming systems.

A recent study found that the array aperiodicity introduced by small random errors in the antenna element placement can be beneficial for the performance of massive multi-user (MU) MIMO communication systems~\cite{Ge_MUMIMO_Aperiodic:2016}. However, to fully understand its advantages, it is necessary to examine and synthesize optimal aperiodic array layouts.
To the best of the author's knowledge, this it the first time that aperiodic array synthesis is proposed and studied in this context.

\begin{figure}[!t]
\centering
    \includegraphics[width=\columnwidth]{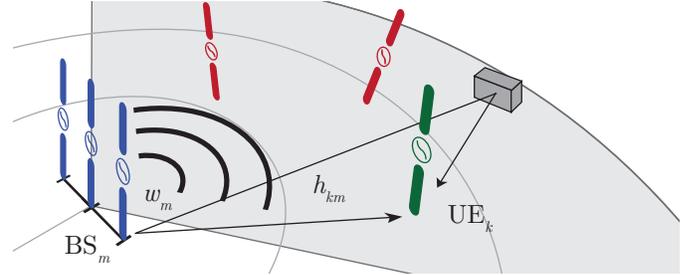}
    \caption{\small Illustration of an $M\times K$ MU-MIMO system in downlink, where $M$ and $K$ are the number of Base Station (BS) antenna elements and User Equipments (UE), respectively.}
    \label{fig:system}
\end{figure}
This paper introduces an innovative deterministic-statistical approach for the synthesis of optimal aperiodic array antennas as base stations for MU-MIMO applications. We first derive the statistical distribution of the element excitations for a dense regular array for a preselected adaptive beamforming algorithm and propagation environment, which is then used as a density taper~\cite{Angeletti_TAPER_LINEAR:2014} to identify a reduced set of optimal array element locations. This paper presents: (i) a mathematical formulation of the proposed approach, and; (ii) numerical results illustrating the effects of different propagation environments, as well as the number of base station antenna elements and users, on the the performance improvements of aperiodic arrays over regular ones. For these comparisons we employ the Zero Forcing adaptive beamformer.

\section{MU-MIMO System Model}\label{System}
In the following subsections we introduce: 1) the MU-MIMO antenna system model for the downlink scenario; 2) the radio propagation channel model; 3) the key communication link performance metrics, employed to evaluate system performance such as the SINR and sum-rate capacity, and; 4) the figures of merit to evaluate power level variations of amplifiers due to adaptive beamforming. Finally, in 5), the uplink duality is briefly outlined.

\subsubsection{Downlink}
Consider the downlink of an $M\times K$ single-cell narrowband MU-MIMO system, as shown in Fig.~\ref{fig:system}.
The BS is equipped with $M$ antennas serving $K$ single-antenna UEs (User Equipments), with $M\ge K$.
Let $\MatVec{x} \in\mathbb{C}^{M\times 1}$ be the signal transmitted from the BS array with normalized power $||\MatVec{x}||^2=1$.
The received signal $\MatVec{y} \in\mathbb{C}^{K\times 1}$ at the $K$ UEs can be expressed as
\begin{equation}\label{eq:y}
\MatVec{y}=\sqrt{\text{SNR}}\MatVec{H}\MatVec{x}+\MatVec{n},
\end{equation}
where $\text{SNR}$ is the average per-user signal-to-noise ratio (SNR), $\MatVec{H}\in\mathbb{C}^{K\times M}$ is the downlink MU-MIMO channel matrix and $\MatVec{n}\in\mathbb{C}^{K\times 1}$ is the additive white Gaussian noise at the users, here assumed having zero-mean and unit variance.

\subsubsection{Channel Model}\label{sec:channel}
The channel matrix $\MatVec{H}$ captures the radio propagation conditions between the antenna ports of the BS and that of the multiple UEs.
In our model, the BS antenna is a linear array of $M$ Huygens sources serving $K$ UEs in a $120^\circ$ cell sector.
Each UE is represented by a linearly polarized antenna with a random orientation and position, see Fig.~\ref{fig:system}.

The MIMO system can be characterized in two extreme propagation environments: (i) the Rich Isotropic MultiPath (RIMP) environment~\cite{Carlober_RIMPRLOS:2010}, with a very large number of random uniformly distributed scattered waves, and; (ii) the Random Line of Sight (RLOS) environment~\cite{RLOS:2013}, with a single direct wave with random properties due to arbitrary position and orientation of the UE.
To characterize these, as well as in-between environments, the problem was modeled by associating to each UE a set of waves (from 1 for RLOS up to 20 for RIMP) with a uniform random distribution in angle of arrival, amplitude, phase and polarization. Note that each wave thus represents either the LOS component or a strong scatterer.
Simulations are repeated $10^6$ times to produce accurate statistics, such as the cumulative probability distribution (CDFs), the average and the variance.


\subsubsection{SINR and Sum Rate Capacity}
We assume perfect channel state information at the transmitter, so that $\MatVec{H}$ is known at the BS. In linear precoding, the transmitted signal can then be expressed as $\MatVec{x}=\sqrt{\beta}\MatVec{W}\MatVec{q}$, 
where $\MatVec{W} \in\mathbb{C}^{M\times K}$  is the precoding matrix, $\MatVec{q} \in\mathbb{C}^{K\times 1}$  is the intended transmitted signal, $\beta=1/\tr(\MatVec{W}\MatVec{W}^\dagger)$  is the power normalization constant, and $^\dagger$ denotes the conjugate transpose operation.

Since we expect an interference-limited scenario, a Zero Forcing (ZF) beamformer is assumed to suppress the interference between UEs. Accordingly, $\MatVec{W}=\MatVec{H}^\dagger (\MatVec{H}\MatVec{H}^\dagger)^{-1}$.

The downlink signal-to-interference-plus-noise ratio (SINR) of the $k$th user can be expressed as~\cite{Sifaou:2017}
\begin{equation}\label{eq:SINR}
\text{SINR}_k=\frac{\beta\text{SNR}|\MatVec{H}_{k,:}\MatVec{W}_{:,k}|^2}{\beta\text{SNR}\sum_{j\ne k}^K{|\MatVec{H}_{k,:}\MatVec{W}_{:,j}|^2}+1}
\end{equation}
where $\MatVec{X}_{i,:}$ and $\MatVec{X}_{:,j}$ indicate the $i$th row and $j$th column of the matrix $\MatVec{X}$, respectively. It is worthwhile to note that Eq~\eqref{eq:SINR} is valid for any pre-coding matrix.


The ergodic sum rate capacity of the total MU-MIMO system is then defined as~\cite{Glazunov:2016}
\begin{equation}\label{eq:SR}
\text{SR}=\sum_{k=1}^{K}{\E[ \log_2{(1+\text{SINR}_k}) ]}.
\end{equation}

\begin{figure}[!t]
	\centering
	\includegraphics[width=.85\columnwidth]{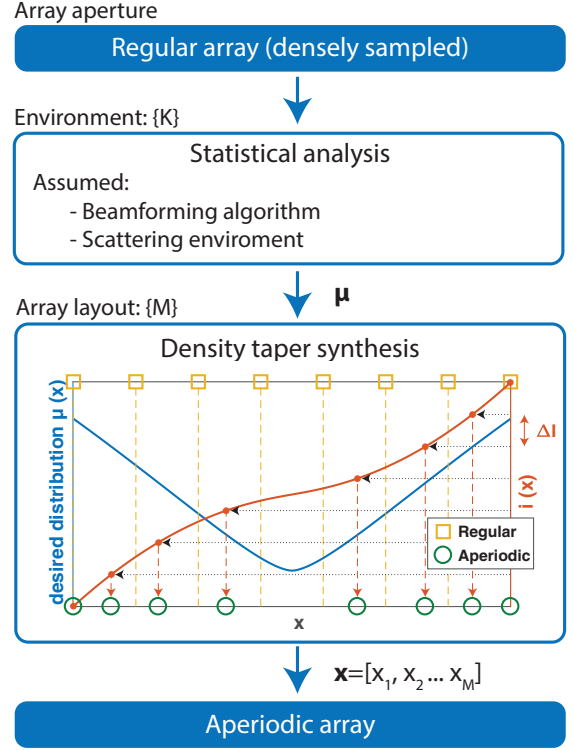}
	\caption{\small Illustration of the proposed dual-step aperiodic array design procedure, as described in Sec.~\ref{sec:synthesis}.}
	\label{fig:design}
\end{figure}

\subsubsection{Amplifier Output Power Levels}
The pre-coding matrix $\MatVec{W}$ is changed adaptively in time as it depends on $\MatVec{H}$, and thus different BS antenna excitation coefficients $\MatVec{x}$ are obtained at every time instant for each UE.
Therefore, to study the statistics of the amplifier output power levels we introduce
\begin{align}\label{eq:mu}
\MatVec{\mu}&=\E\left[\left|\sum_{k=1}^{K}{\MatVec{W}_{:,k}}\right|^2\right];		&\MatVec{\sigma}^2&=\Var\left[\left|\sum_{k=1}^{K}{\MatVec{W}_{:,k}}\right|^2\right]
\end{align}
where $\MatVec{\mu}$ and $\MatVec{\sigma}$ are the vectors describing the average power and its variance for each PA (power amplifier), when transmitting to all users at the same time.
When transmitting to one user only at a time instead, the above expressions reduce to $\MatVec{\mu}_k=\E[|\MatVec{W}_{:,k}|^2]$ and $\MatVec{\sigma}_k^2=\Var[|\MatVec{W}_{:,k}|^2]$, however identical distribution are found in both scenarios.
We define the power spread as
\begin{align}\label{PS}
\text{PS}=\max(\MatVec{\mu}+\MatVec{\sigma}^2)/\min(\MatVec{\mu}-\MatVec{\sigma}^2)
\end{align}
where PS=0dB represents the ideal constant and uniform power level for all PAs, while larger values indicate a stronger variation of the excitations across the array and/or channel realizations.
In practice, in the downlink scenario, PAs are deployed at each of the BS antenna ports to provide the desired radiated power.
Solid state PAs are designed for a fixed maximum output power and typically operate most efficiently at saturation~\cite{MIMO_AMPL_Persson:2014}.
It is thus desirable to have all power amplifiers operating at a uniform power level (thus lower PS values), but this is difficult to reach due to the high degree of adaptivity of MIMO systems. 

\subsubsection{Uplink}
Similar expressions to \eqref{eq:y} apply to the uplink scenario, where $\MatVec{y}\in\mathbb{C}^{M\times 1}$ is the received signal at the BS and $\MatVec{x} \in\mathbb{C}^{K\times 1}$ is the signal transmitted from the UEs. The decoded signal can instead be expressed as $\MatVec{\tilde y}=\MatVec{D}\MatVec{y}$, where $\MatVec{D}$ is the decoding matrix. The $\text{SINR}$ expression \eqref{eq:SINR} is modified by the asymmetry of the link, however the SR expression \eqref{eq:SR} is the same as the one for the uplink. We refer, e.g., to~\cite{Ngo_MIMO_Massive:2015} for a more complete formulation.

\section{Design Methodology}\label{Method}
In the following we propose a novel design approach to the synthesis of aperiodic arrays tailored to the MU-MIMO type scenario.

\setcounter{subsubsection}{0}
\subsubsection{Aperiodic array design}\label{sec:synthesis}


The newly proposed synthesis method is based on a combined statistical analysis and a density taper approach, where the knowledge of the statistical distribution of the excitations is used to determine the optimal array layout, see Fig.~\ref{fig:design}.
Firstly, we densely sample the desired aperture with a regular array of Huygens sources to represent a generic aperture field distribution. Then, we simulate the dense array in the desired propagation environment, as described in Section~\ref{sec:channel}, and compute the resulting average powers $\MatVec{\mu}$ [see~\eqref{eq:mu}] of these array elements. This will form our reference power distribution. Subsequently, the aperiodic layout is synthesized through the density taper approach~\cite{Angeletti_TAPER_LINEAR:2014}, i.e., elements are located with a density proportional to the reference power distribution. Mathematically, the antenna positions are obtained as
\begin{equation}
x_m=i((m-1)\Delta I)^{-1} \quad  m=1,\ldots M
\end{equation}
where $i(x)=\int_0^x {\mu(\tau) \text{d}\tau}$ is the auxiliary cumulative distribution derived from $\mu(x):[0,X_\text{max}]$, $\Delta I=i(X_\text{max})/(M-1)$ is its equipartition and $i^{-1}$ denotes the inverse operation.
As shown in Fig.~\ref{fig:design}, starting from the reference distribution $\mu(x)$, the antenna positions are easily found as the intersection points between $i(x)$ and its equipartitions.

\subsubsection{Aperiodic array gains}\label{sec:figuresofmerit}
For the assessment of the performance of the aperiodic array, all results are presented in comparison with the respective uniform array, having the same aperture, antenna type and total number of antenna elements. Besides individual performance curves, relative gain curves are plotted to indicate the improvement of the aperiodic over the regular array. For instance, in terms of the link quality, the SINR Gain (SINRG) is introduced,
\begin{equation}
\text{SINRG}=[\text{SINR}^\text{aperiodic}/\text{SINR}^\text{regular}]^{\text{SNR}=0\text{dB}}_{\text{CDF=5\%}}
\end{equation}
as the SINR difference between the two arrays for an SNR of 0dB evaluated at the 5\% level of the CDF (i.e. for 95\% of the users).
The same SINRG is obtained for each of the UE streams due to the random nature of the scenario.

Similarly, regarding the amplifier output powers, the Power Spread Compression (PSC) is defined as the power spread difference between the two arrays, i.e.,
\begin{equation}
\text{PSC}=[\text{PS}^\text{regular}/\text{PS}^\text{aperiodic}].
\end{equation}
Both the SINRG and the PSC are positive (in dB) when the aperiodic array outperforms the regular one.

\section{Results}\label{Results}
\begin{figure}[!t]
	\centering
	\includegraphics[width=\columnwidth]{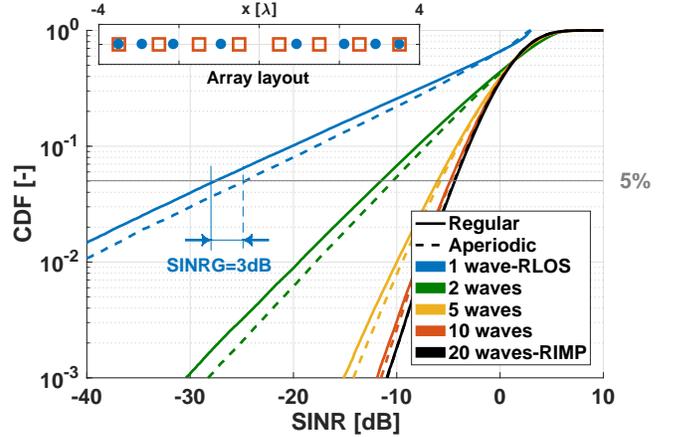}
	\caption{\small Uplink Cumulative Distribution Functions (CDFs) of SINR for an $8\times 2$ MU-MIMO system operating in different propagation environments, which are modeled by a number of incoming plane waves, one per UE. For all propagation cases SNR=0dB and the results are shown for the regular (solid lines) and optimized aperiodic (dashed lines) array antennas (see the layout inset). }
	\label{fig:environment}
\end{figure}

\begin{figure}[!t]
	\centering
	\includegraphics[width=\columnwidth]{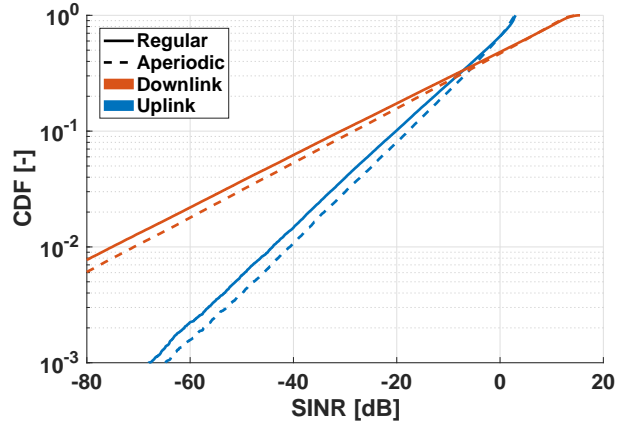}
	\caption{\small Uplink and Downlink CDFs of SINRs at SNR=0dB for an $8\times 2$ MU-MIMO system, in the Random Line-Of-Sight (RLOS) propagation environment.}
	\label{fig:uplinkdownlink}
\end{figure}

In the following we discuss: 1) the effect of the propagation environment, 2) the uplink/downlink duality, 3) the impact of the aperiodic layout on the SINR and the SR, as well as 4) on the amplifier output powers.
Different systems are compared, whose sizes range from $8\times2$ up to $16\times8$ ($M\times K$), and where the solid and dashed curves are for the regular and aperiodic array case, respectively. In all cases the array aperture is $(M-1)\lambda$. The SINR CDF, Eq.~\eqref{eq:SINR}, and the SR, Eq.~\eqref{eq:SR}, are used to discuss the link reliability and the capacity, respectively.
Finally, in 5), large scale massive MIMO systems are examined in terms of the relative gains as defined in Sec.~\ref{sec:figuresofmerit}.

\setcounter{subsubsection}{0}
\subsubsection{LOS- to RIMP-dominated environments}\label{sec:los}
We study scattering effects by changing the number of plane waves associated with each UE (see Sec.~\ref{sec:channel}).
Fig.~\ref{fig:environment} compares the SINR CDFs for an $8\times2$ aperiodic and regular array, when moving from an RLOS- to RIMP-dominated environment.
Note that in all SINR CDF plots, rightmost and steepest curves are preferred since then higher SINR values are more probable.
For RLOS (1 random wave per UE), the aperiodic array offers the largest gain (SINRG=3dB), while it progressively reduces for increasing scattering until the two curves overlap for the RIMP environment (10-20 random waves per UE); Thus the aperiodic array always exhibits superior or identical SINRG performance.
Evidently, the RLOS environment is the most favorable propagation condition for aperiodic arrays and therefore considered in the remainder of the paper.

\subsubsection{Uplink and Downlink}
\begin{figure}[!t]
	\centering
	\includegraphics[width=\columnwidth]{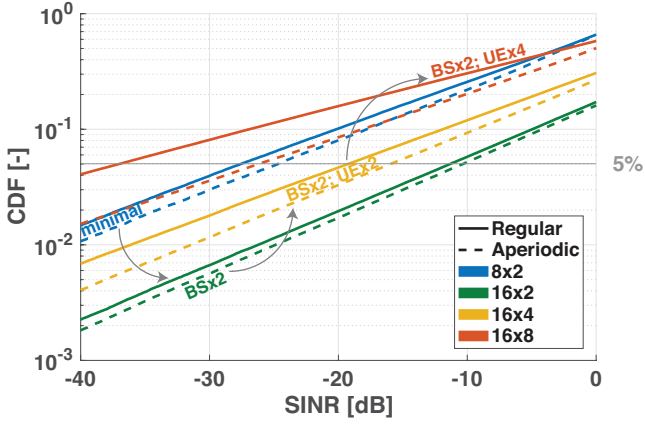}
	\caption{\small Uplink CDFs of SINR at SNR=0dB for different MU-MIMO system sizes, in the RLOS environment.}
	\label{fig:SINR}
\end{figure}
\begin{figure}[!t]
	\centering
	\includegraphics[width=\columnwidth]{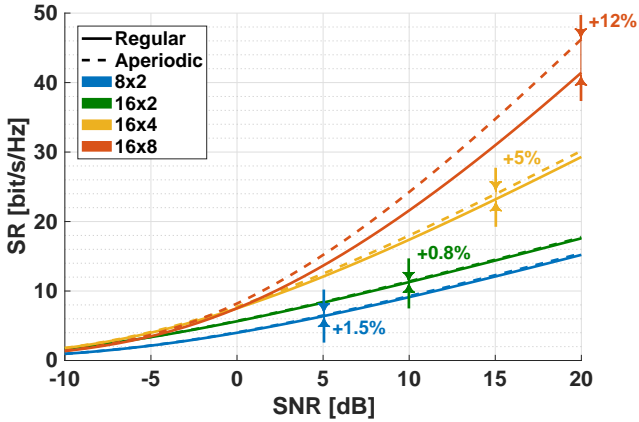}
	\caption{\small Uplink Sum Rate (SR)  for different MU-MIMO system sizes, in the RLOS environment.}
	\label{fig:SR}
\end{figure}
Similar conclusions are valid for downlink and uplink when it comes to the SINR gain.
To show this, the CDFs of the aperiodic and regular arrays in the two scenarios are plotted in Fig.~\ref{fig:uplinkdownlink}.
Although the distributions are different due to the different SINR expressions, the aperiodic SINR gain is evidently present and of comparable value in both cases.
Regarding the antenna excitations, identical power profiles $\mu(x)$ are found, which is due to the pre- and de-coding matrices, one being the Hermitian of the other.
However, the efficiencies of the low-noise amplifiers in the receive mode are not as important as those for the PAs in the transmit case.
Note that, in the uplink scenario: (i) the CDF shows directly the required transmitted power for a minimum SINR to a certain user percentile, and; (ii), the SINR CDF curves are linearly dependent on the SNR~\cite{Chen:2014}.
We thus choose to show the rest of the SINR results only for the uplink scenario and for one SNR only.

\subsubsection{SINR and Capacity}
\begin{figure}[!t]
	\centering
	\includegraphics[width=\columnwidth]{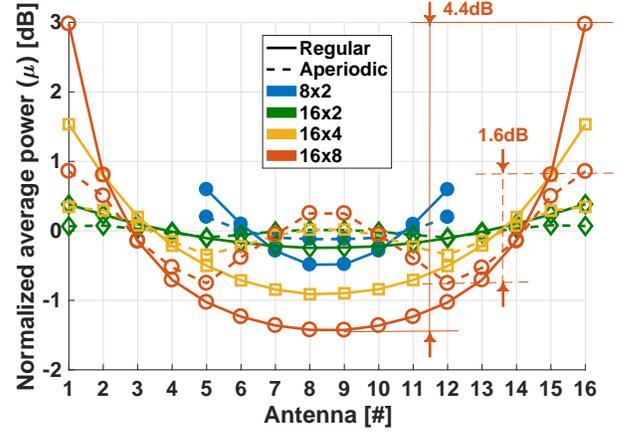}
	\caption{\small Amplifier's normalized average output power levels for different MU-MIMO system sizes, in the RLOS environment.}
	\label{fig:average}
\end{figure}
\begin{figure}[!t]
	\centering
	\includegraphics[width=\columnwidth]{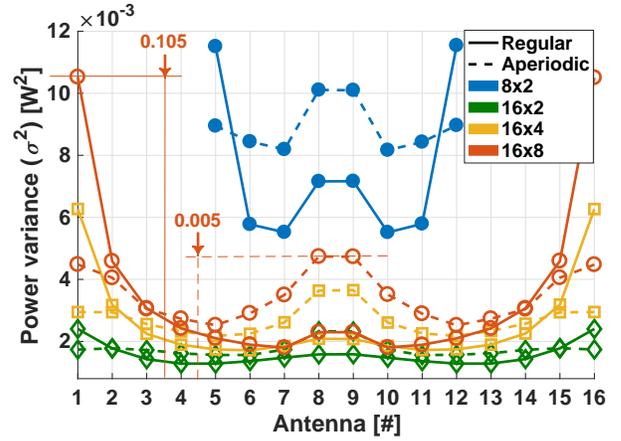}
	\caption{\small Amplifier's power variance for MU-MIMO of different system sizes, in RLOS.}
	\label{fig:variance}
\end{figure}

As shown above, aperiodic arrays can improve the SINR ratio, and thus the capacity too.
In Fig.~\ref{fig:SINR} the SINR CDFs for the aperiodic and regular arrays of different system sizes are compared.
Starting from the $8\times 2$ system (SINRG=3dB), if we double the number of BS antennas, but keep the number of UEs the same, the SINR CDF improves (i.e. moves to the right) as expected. However, the SINRG decreases to 1dB.
If, on the other hand, the number of UEs is also doubled, and thus the ratio of the number of BS-to-UE antennas is kept constant, not only the CDFs improve but also the SINRG increases to 3.5dB.
Finally, if the UEs are further doubled, the SINRG exceeds 10dB: note how the $16\times 8$ aperiodic array provides approximately the same per-user SINR of the $8\times 2$ array, where the regular array case would instead lose 10dB.

Fig.~\ref{fig:SR} shows the SR capacities corresponding to the same array configurations as shown in Fig.~\ref{fig:SINR}.
The system capacity increases with both the number of BS antennas and the number of UE, with the $16\times 8$ aperiodic array having a 12\% rate increase over the regular.
That is, both the user's 5-percentile SINR improves (which condition the link budget), as well as the capacity in more crowded scenarios.

\subsubsection{Amplifier power}
Fig.~\ref{fig:average} shows the normalized average antenna output powers of the regular and aperiodic arrays that were considered above. For the regular array, increasing the BS antennas and the UE-to-BS ratio exacerbates the power unbalance between the edge and central elements.
The aperiodic array, on the other hand, exhibits a more uniform average port power among the elements: the $16\times 8$ aperiodic array has a 2.8dB tapering reduction.

A similar trend is observed for the variance in antenna port powers as shown in Fig.~\ref{fig:variance}.
Here too, the highest variance is associated with the edge elements, irrespective of the system size. The aperiodic array reduces the variance, albeit to a lesser extent for the central elements; hence, aperiodicity is mostly beneficial in ensuring a uniform average power allocation.

\subsubsection{Massive MU-MIMO}\label{sec:massive}
\begin{figure}[!t]
	\centering
	\includegraphics[width=\columnwidth]{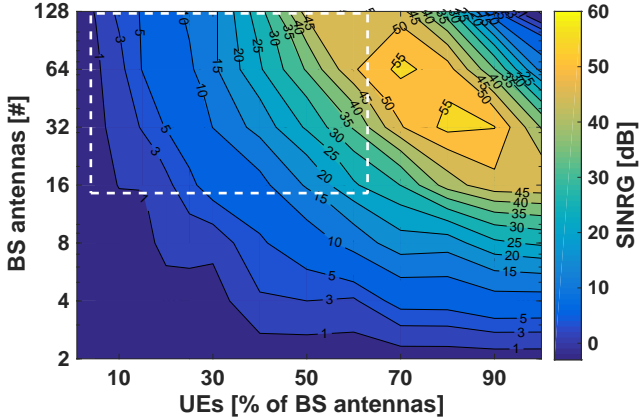}
	\caption{\small Aperiodic array gain as function of BS antennas and cell crowdedness (number of UE as \% of BS antennas).}
	\label{fig:PGainvsUEsvsBSs}
\end{figure}

\begin{figure}[!t]
	\centering
	\includegraphics[width=\columnwidth]{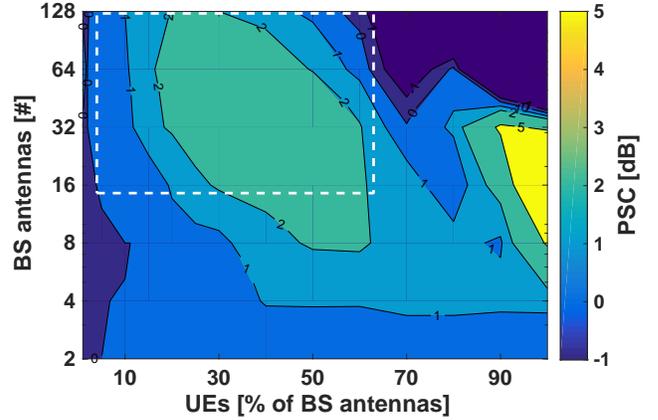}
	\caption{\small Power spread compression as function of BS antennas and cell crowdedness (number of UE as \% of BS antennas).}
	\label{fig:CompvsUEsvsBSs}
\end{figure}

The link quality (SINR) and the power spread of PAs are strongly dependent on the system size. It is therefore relevant to study the impact of the massive MIMO architecture on the SINRG and the PSC figures-of-merit.
Fig.~\ref{fig:PGainvsUEsvsBSs} shows the SINRG as a function of the number of BS antennas and cell crowdedness, i.e., the number of UEs as a percentage of the BS antennas ($K/M*100$).
In Fig.~\ref{fig:CompvsUEsvsBSs} the PSC is shown as well.
In general, the SINRG can be very substantial, particularly in crowded cells and for a large number of BS antennas. The PSC, on the other hand, demonstrates more moderate gains.
As an example: a moderately large aperiodic system of 64 BS antennas would experience an SINR increase of 3dB at 10\% UEs, and more than 15dB at 30\%, while having a PSC between 1 to 3dB.

\section{Conclusions}\label{Conclusions}
In this manuscript we have investigated the advantages of aperiodic arrays for MU-MIMO applications.
We have introduced a simple aperiodic design method based on a hybrid statistical-density tapering approach.
We have then considered the effects on: (i) the link performance, and; (ii) the amplifier power spread with respect to classical regular arrays.

Results show that aperiodic MU-MIMO arrays provide the largest gains in line-of-sight dominated environments.
This reduces for increasing degree of scattering, however they are never inferior to the regular arrays.
This can be concluded in both the up and downlink case.
Aperiodic arrays are shown to be beneficial to the link performance, both to the users' 5 percentile SINR as well as the sum rate capacity.
Moreover, the aperiodic layout improves the amplifier´s efficiency owing to a more uniform average power among the antenna's power amplifiers.
Even for a relatively small $16\times 8$ MU-MIMO system, it is possible to achieve a 10dB power budget improvement, a 12\% capacity increase and a 3dB amplifier tapering reduction.
Finally, it is shown that larger and more crowded MU-MIMO systems benefit most by the array aperiodicity.
Results show that aperiodic arrays can provide a substantial gain in the link quality and capacity, especially in scenarios with large interference.

\bibliographystyle{IEEEtran}
\bibliography{IEEEabrv,AperiodicMIMO}

\end{document}